\documentclass[preprint]{vgtc}               

\graphicspath{{figures/}{pictures/}{images/}{./}} 

\usepackage{times}                     

\usepackage{tabu}                      
\usepackage{booktabs}                  
\usepackage{lipsum}                    
\usepackage{mwe}                       

\usepackage{mathptmx}                  
\usepackage{tcolorbox}
\usepackage{xcolor}
\usepackage{listings}
\usepackage{enumitem}
\usepackage[table]{xcolor}
\usepackage{amsmath,amsfonts}
\usepackage{graphicx}  
\usepackage{array, multirow, booktabs}
\usepackage[nocompress, sort]{cite}
\usepackage{tikz}
\usepackage[normalem]{ulem}

\definecolor{overview_red}{HTML}{FF9797}
\definecolor{overview_blue}{HTML}{A0C8FF}
\definecolor{overview_green}{HTML}{95DC9F}
\definecolor{paradigm_gray}{HTML}{E0E0E0}

\definecolor{lightred}{RGB}{255, 175, 177}
\definecolor{lightorange}{RGB}{255, 215, 174}
\definecolor{lightyellow}{RGB}{255, 250, 194}

\definecolor{boxplot_gray}{HTML}{B1B1B1}
\definecolor{boxplot_green}{HTML}{A8DEAF}
\definecolor{boxplot_red}{HTML}{F497A3}

\definecolor{kim_green}{HTML}{4CBB17}
\definecolor{yalong_purple}{HTML}{9300FF}
\definecolor{ari_red}{HTML}{EC1313}

\definecolor{pending_user_study_res}{HTML}{2171DF}

\newcommand{\para}[1]{\vspace{0.35em}\noindent\normalsize\textbf{#1.}\xspace}
\newcommand{\parac}[1]{\vspace{0.35em}\noindent\normalsize\textbf{#1:}\xspace}

\newcommand{\sys}[0]{SpeechLess\xspace}

\newcommand{\suppl}[0]{Supplementary Material\xspace}

\newcommand{\fullq}[0]{\textit{Full}\xspace}
\newcommand{\partq}[0]{\textit{Partial}\xspace}
\newcommand{\silentq}[0]{\textit{Zero}\xspace}

\newcommand{\dc}[2][1em]{%
  \raisebox{-0.5\dp\strutbox}{\includegraphics[height=#1]{#2}}%
}

\makeatletter
\newcommand{\recaptitle}[1]{%
  \noindent{\reset@font\sffamily\normalsize\vgtc@sectionfont\itshape #1.}%
}
\makeatother

\makeatletter
\newcommand{\cleansubsec}[1]{%
  \subsubsection*{#1}%
  \phantomsection 
  \addcontentsline{toc}{subsubsection}{#1}%
}
\makeatother

\onlineid{1402}

\vgtccategory{System}

\vgtcinsertpkg

\usepackage{orcidlink}

\preprinttext{%
  \ifodd\value{page}%
    \parbox{0.9\textwidth}{%
      © 2026 IEEE. This is the author's version of the article that will appear at the IEEE Conference on Virtual Reality and 3D User Interfaces (IEEE VR). The final version of this record is available at: \href{https://doi.org/10.1109/VR67842.2026.00044}{\textcolor{blue}{\textit{10.1109/VR67842.2026.00044}}}
    }%
  \fi
}

\title{SpeechLess: Micro-utterance with Personalized Spatial Memory-aware Assistant in Everyday Augmented Reality}

\author{
  \begin{tabular}{@{}c@{\hspace{5em}}c@{\hspace{5em}}c@{}}
    Yoonsang Kim\thanks{e-mail:yoonsakim@cs.stonybrook.edu} &
    Devshree Jadeja\thanks{e-mail:devshreehardik.jadeja@stonybrook.edu} &
    Divyansh Pradhan\thanks{e-mail:divyansh.pradhan@stonybrook.edu}\\
    \scriptsize Stony Brook University &
    \scriptsize Stony Brook University &
    \scriptsize Stony Brook University
  \end{tabular}
  \\[0.75em]
  \begin{tabular}{@{}c@{\hspace{5em}}c@{}}
    Yalong Yang\thanks{e-mail:yalong.yang@gatech.edu} &
    Arie E. Kaufman\thanks{e-mail:ari@cs.stonybrook.edu}\\
    \scriptsize Georgia Institute of Technology &
    \scriptsize Stony Brook University
  \end{tabular}
}

\teaser{
  \centering
  \vspace{-5mm}
  \includegraphics[width=\linewidth]{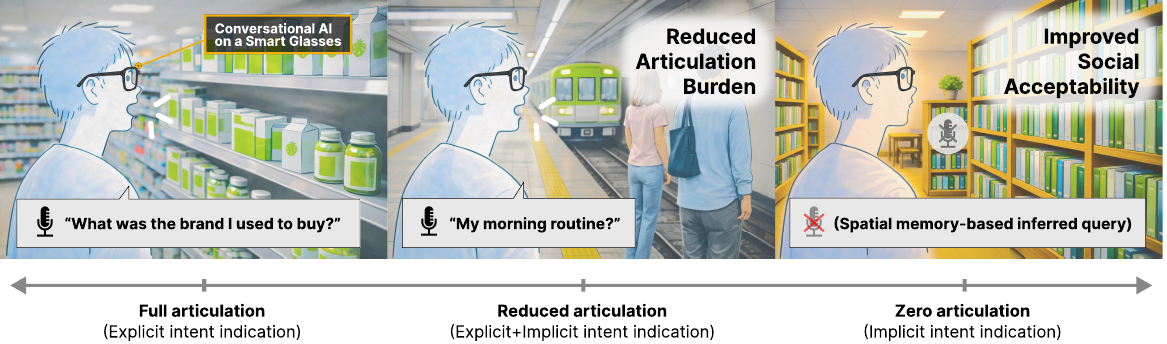}
  \vspace{-5mm}
  \caption{
  Concept illustration of \sys. Users' queries are stored in the form of a ``Spatial memory'' along with their spatiotemporal, activity, and intent contexts. Later, under the similar/same routine context, these memories can be unobtrusively recalled with micro/zero-utterance, reducing the overhead of identical query repetition or minimal exposure of query contents to bystanders.
  }
  \label{fig:teaser}
}

\abstract{
Speaking aloud to a wearable AR assistant in public can be socially awkward, and re-articulating the same requests every day creates unnecessary effort. We present \sys, a wearable AR assistant that introduces a speech-based intent granularity control paradigm grounded in personalized spatial memory. \sys helps users ``\textit{speak less},'' while still obtaining the information they need, and supports gradual explicitation of intent when more complex expression is required. \sys binds prior interactions to multimodal personal context--space, time, activity, and referents--to form spatial memories, and leverages them to extrapolate missing intent dimensions from under-specified user queries. This enables users to dynamically adjust how explicitly they express their informational needs, from full-utterance to micro/zero-utterance interaction. We motivate our design through a week-long formative study using a commercial smart glasses platform, revealing discomfort with public voice use, frustration with repetitive speech, and hardware constraints. Building on these insights, we design \sys, and evaluate it through controlled lab and in-the-wild studies. Our results indicate that regulated speech-based interaction, can improve everyday information access, reduce articulation effort, and support socially acceptable use without substantially degrading perceived usability or intent resolution accuracy across diverse everyday environments.
} 


\begin{document}


\firstsection{Introduction}

\maketitle

Augmented Reality (AR) provides a powerful mechanism for embedding digital information within the physical world, augmenting human perception and interaction~\cite{cai2025aiget}. By situating digital content in the user's surroundings, AR enables access to information that is inherently tied to physical context~\cite{lee2023design}. Recent advances in Large Language Models (LLMs) further extend this capability by supporting conversational interfaces that interpret natural language and respond in context. The convergence of AR and LLMs has opened new opportunities for intelligent assistants that not only situate content in the physical realm, but also reason about user intent through interaction, and enable more fluid and context-aware assistance~\cite{dogan2024augmented, lee2024gazepointar, shenencode24}. The introduction of wearable AR extends this vision further, by embedding AI assistance into an always-available, first-person view form factor~\cite{yang2025contextagent, paruchuri2025egotrigger}. 

Wearable AR assistance can operate continuously within users' daily routines, where rich embodied context such as location, activity, and environment, naturally accompanies each moment of use. This continuous nature of wearable AR's context enables the system to better identify recurring everyday usage patterns and proactively anticipate the user's needs~\cite{lee2025sensible, liu2024proactive, pu2025promemassist, meurisch2020exploring, li2025satori}. While existing wearable AR systems support proactive assistance, they largely provide limited mechanisms for users to regulate how their intent is specified or disambiguated when anticipatory inference is uncertain, socially inappropriate, or misaligned with moment-to-moment needs (e.g., binary confirmation, multiple-choice selection)~\cite{xu2023xair}. To this end, wearable AR requires a new ``\textit{interaction paradigm}'' that complements anticipatory inference, with a mechanism allowing context-rich expression of user intent--enabling users to dynamically specify or refine their needs while balancing convenience and reliability~\cite{davari2024towards, rajaram2025gesture, grubert2016towards}.

We introduce \textbf{\textit{\sys}}, a wearable AR assistant that instantiates a new interaction paradigm for everyday use, while ensuring both convenience and robustness. We couple (1) long-term personalized spatial memory to represent recurring user activities, and (2) speech as a mechanism for dynamically adjusting the granularity of activity intent specification, depending on social context and situational demands (e.g., \cref{fig:teaser}--adjusting query style by social context). We design our proof-of-concept system informed by a 7-day formative study using a commercially available, conversational wearable AI assistant (Meta RayBan), which revealed recurring challenges with full speech-based interaction in public settings, frustration arising from repetitive verbal queries, and practical hardware limitations. To assess whether \sys addresses these concerns, we evaluate the system across both public and private usage contexts, including crowded everyday environments and controlled laboratory settings. We examine how effectively \sys supports intent resolution, reduces articulation burden, and maintains usability, under varying social and situational conditions.

By grounding information access in recurring everyday contexts, \sys advances conversational assistance in wearable AR, toward a more context-aware and adaptive agent that allows users to regulate the granularity of intent expression (from micro to fully articulated verbal queries) based on social and situational constraints, supporting low-effort, socially acceptable interaction in daily activities. Our core contributions are as follows:
\begin{itemize}
\item \parac{Interaction paradigm for everyday wearable AR}
The proposal of an interaction paradigm that allows users to control query intent granularity through speech, adapting to their needs, social, and situational context.
\vspace{-2.5mm}
\item \parac{Personal spatial memory representation design}
A spatially grounded memory representation binding digital information with personal multimodal context--speech, space, time, actions, and referents--in the real-world for context-aware intent resolution.
\vspace{-2.5mm}
\item \parac{Practical challenges on current smart wearables}
Empirical findings from a longitudinal in-the-wild study (with Meta RayBan), highlighting the current limitations of a speech-based smart wearable device in public/private spaces.
\vspace{-2.5mm}
\item \parac{Demonstration and validation}
Validation of improved convenience and social acceptability, while maintaining accurate intent resolution under controlled visualization and socially constrained conditions (in-lab, crowds).
\end{itemize}

\section{Related Work}
\recaptitle{AR at the Corner of Everyday Life}
The growing integration of AR into daily life is reflected in popular applications such as Pokemon Go and Google Maps Live View~\cite{rau2022supporting, tran2025wearable}. Prior works have explored how AR can support everyday use by situating information directly within physical environments. XR-Objects~\cite{dogan2024augmented} extends the role of AR in everyday environments by retrieving knowledge with mobile AR via LLM. Others explore how AR enhances daily activities, including human memory recall and daily task support~\cite{humanmemory, bressa2022data, fang2025mirai, remberingmemory, shenencode24, omar2025}. These works show the potential for AR to be an integral part of our everyday life. Building on the notion of ``everyday AR'', we focus on capturing daily routine AR interactions to model long-term personal interests and their context.

\vspace{2mm}
\recaptitle{Context-aware AI Assistant and LLM}
LLMs are employed to contextualize multimodal input into conversational responses, enabling interpretation of intent and task-guidance~\cite{stover2024taggar, zhu2025agentar}. Prior works leverage LLM for accessibility or context-aware question answering~\cite{chang2024worldscribe, hu2025vision}, or adopt LLM for proactive agents that anticipate user needs~\cite{li2025satori, han2025towards, zhang2024ask}. Sensible Agent shows how agent proactivity (system-triggered) and context-aware modality switching, can be used for enhancing socially acceptable AI~\cite{lee2025sensible}. Personalized contextual responses through user embeddings or situational cues, are also explored~\cite{khan2024situational,userllm, arakawa2024prism, fung2025embodied, wang2025if, long2025seeing}. These show how LLMs can ground multimodal inputs with context-aware responses. We leverage LLM to fuse user contexts (spatiotemporal, activity, and object) to infer intentions of user-spoken queries (user-initiated)~\cite{chu2025intention, kundu2025pilar}.

\vspace{2mm}
\recaptitle{Binding Information with Contextual Cues}
Binding digital information to contextual cues--conversational, spatial, or visual--has been a longstanding goal in ubiquitous computing~\cite{perera2013context, li2024omniquery}. OmniActions predicts user intents based on multi-source pattern learning~\cite{omniactions}. Memoro demonstrates how a lightweight, audio-based assistant can log everyday conversations and support remembrance of prior queries through conversational replay~\cite{zulfikar2024memoro}. Beyond conversational traces, prior works show the binding of AR content to physical contexts~\cite{dogan2024augmented, kim2025explainable, buschel2021miria, liu2024investigating, assor2023handling}. Encode-Store-Retrieve illustrates how egocentric perception can be encoded into language for episodic memory augmentation~\cite{shenencode24} and psychological studies of context-dependent memory highlight how environmental cues support recall~\cite{satriadi2023context,easton1995object}. Other works leverage contextual and prior knowledge to support user assistance~\cite{yang2025thinking, akbar2025rag}. Our work extends this trajectory by using spatially anchored personal memory not only for a reminder mechanism~\cite{pu2025promemassist, remberingmemory}, but also for filling missing implicit intent gaps in a spoken query (partially complete query), and leveraging it for convenience and unobtrusiveness in daily settings.

\vspace{2mm}
\recaptitle{Use of Wearable AR in Public}
Deploying an AI assistant on a wearable XR device such as smart glasses or an HMD can raise challenges of social comfort and privacy~\cite{rajaram2025exploring, lee2025sensible, bajorunaite2023reality, bajorunaite2021virtual}. Prior studies report that speech-based interaction in public can feel awkward, intrusive, or socially stigmatizing, while embodied gestural input may be impractical during mobility, or in public contexts~\cite{lu2023wild,tran2025wearable,cheng2025augmented, rajaram2025gesture}. Research on privacy-preserving AR highlights how bystander exposure and unintended disclosure constrain real-world deployments~\cite{corbett2023bystandar, bukhari2025rethinking, erebus, kim2021design}. Glanceable AR emphasizes the need for lightweight, subtle, and socially unobtrusive modalities that reduce interaction effort and visible exposure~\cite{lu2021evaluating}. Building on these insights, our work aims to design an interaction mechanism  that minimizes audible articulation, improving social acceptability and protecting privacy during everyday public routines~\cite{lv2024aria, engel2023project}.
\newline
\vspace{-2mm}

\begin{figure}[!t]
    \centering
    \includegraphics[width=\linewidth]{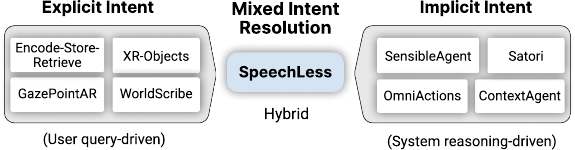}
    \vspace{-6.75mm}
    \caption{\sys comprehends user intents using the benefits of both user-driven (explicit task prompting) and system-driven (proactive intent inference), intent resolution mechanisms.
    }
    \label{fig:overview}
\vspace{-6.25mm}
\end{figure}

As illustrated in \cref{fig:overview}, \sys takes a \textit{hybrid approach} to intent resolution, combining \textit{user-driven explicit} intent specification (e.g., XR-Objects~\cite{dogan2024augmented}, GazePointAR~\cite{lee2024gazepointar}) and \textit{system-driven implicit} intent extrapolation (e.g., Sensible Agent~\cite{lee2025sensible}, OmniActions~\cite{omniactions}). Users dynamically determine how much intent detail to convey through an expressive and natural interaction mode--\textit{\uline{\textbf{speech}}}~\cite{rajaram2025gesture}, while the system proactively extrapolates missing intent dimensions based on the user's personalized spatial context.

\vspace{-1mm}
\section{Design Rationale}
We outline the motivation underlying our design, grounded in the need for a revised interaction paradigm for everyday wearable AR that leverages contextual intelligence to operate under social and situational constraints. We share the insights of our longitudinal formative study (\cref{subsec:formative_study}), collected with a state-of-the-art wearable AI device (Meta RayBan). Then, we introduce the principles that guided the design of \sys (\cref{subsec:design_goals}).

\subsection{Identifying the Practical Challenges of Wearable AI}
\label{subsec:formative_study}
To investigate the practical challenges of current wearable AI assistants, before extending them to AR, we conducted a seven-day formative study (N=8) using a commercially available wearable AI, Meta RayBan glasses. The aim behind the multi-day study was to examine how the devices fit into the participants in realistic daily routines across private and public environments. We focused on three research questions: \textbf{(Q1)} \textit{What unique patterns does wearable smart glasses exhibit, compared to smartphones}; \textbf{(Q2)} \textit{How do individuals perceive and manage themselves in public spaces while wearing the device} (when such form factor is not fully established in people's perception); and \textbf{(Q3)} \textit{How do users feel about speech-based communication with the wearable AI assistant.} 

Participants were recruited from diverse pool of professional backgrounds--researchers, software engineers, and students, and were instructed to wear the device for at least six hours a day (not consecutive), in private and public environments. They were guided to log their daily experiences, and provide qualitative feedback via semi-structured interviews, upon the completion of their participation. Feedback were analyzed using an inductive thematic analysis approach. Three co-authors independently reviewed, and grouped recurring challenges into high-level themes through discussion. Below, we share the coded themes:

\begin{figure}[!t]
    \centering
    \includegraphics[width=\linewidth]{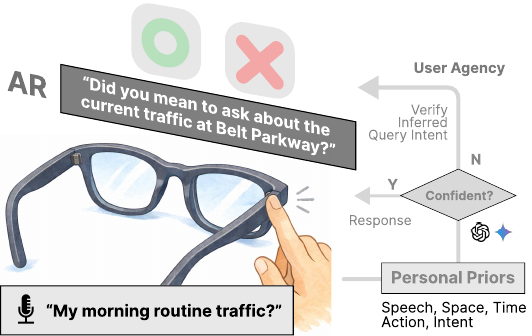}
    \vspace{-6.5mm}
    \caption{Conceptual illustration of device-to-user interface mapping. A minimalistic, hardware-based interface is preferred for the wearable; Human-in-the-loop for a transparent, trustworthy response.}
    \label{fig:confirmation_interface}
\vspace{-5.5mm}
\end{figure}

\vspace{2mm}
\recaptitle{Constraints from Social Norms}
Verbal interaction was perceived as socially inappropriate in certain spaces such as libraries, classrooms (during lecture), or office settings. Participants avoided speaking aloud even when they needed the assistance (N=9). One participant described the dilemma: ``\textit{I wanted to look something up during class, but I just couldn't without disturbing everyone}''. Hesitation due to workplace norms was reported also (N=3). These reflect a strong influence of situational norms on participants willingness to engage with the device.

\vspace{2mm}
\recaptitle{Frustration in Articulation Burden}
Participants expressed frustration with the repetitive invocation of the wake word (``Hey Meta''), describing it as cumbersome (N=6). Several resorted to hardware-based interaction (a physical tap on the side of the glasses) instead. Furthermore, participants reported cases where they were reluctant to use speech out loud for every question (N=5). These reveal the need for a more efficient mechanism to minimize the query effort, and for robustness in assistance triggering.

\vspace{2mm}
\recaptitle{Limited Diverse Use Cases in Daily Routine}
Participants reported that their daily routines left little room for novel interactions with the assistant. ``\textit{Didn't have anything new to ask everyday}''. Many described repetitive schedules involving work-to-home commutes, where they found limited opportunities to ask new meaningful questions beyond their routine (e.g., weather, traffic, bus schedules) (N=6). Several participants described a lack of use cases when the same task could be achieved with their smartphone (N=2).

\vspace{2mm}
\recaptitle{Social Awkwardness and Privacy Concerns}
Participants described heightened public attention while using the glasses in public. Blinking lights intended as a visual indicator for assistants' listening mode, paradoxically drew more attention. ``\textit{People kept staring when the light blinked, even when I was not recording them, or at least I felt that way}''. Privacy-related concerns were reported where bystanders misjudged them for recording (N=5), or assumed they were seeking human assistance when users spoke aloud alone (N=1). Other participants noted that the uncommon form factor (compared to mobile devices) and verbal queries, drew more attention than desired, making them hesitant to initiate assistance (N=6).

\vspace{2mm}
\recaptitle{Hardware-induced Constraints}
The physical and system constraints of the wearable device were pointed out. Short battery life--lasting less than 30 minutes during consecutive video recording (N=1), and paired smartphone overheating during data synchronization (N=2). Others brought up the limitation of recordings resolution, connection reliability, and form factor inconvenience (N=4).


\begin{figure*}[!t]
    \centering
    \includegraphics[width=\linewidth]{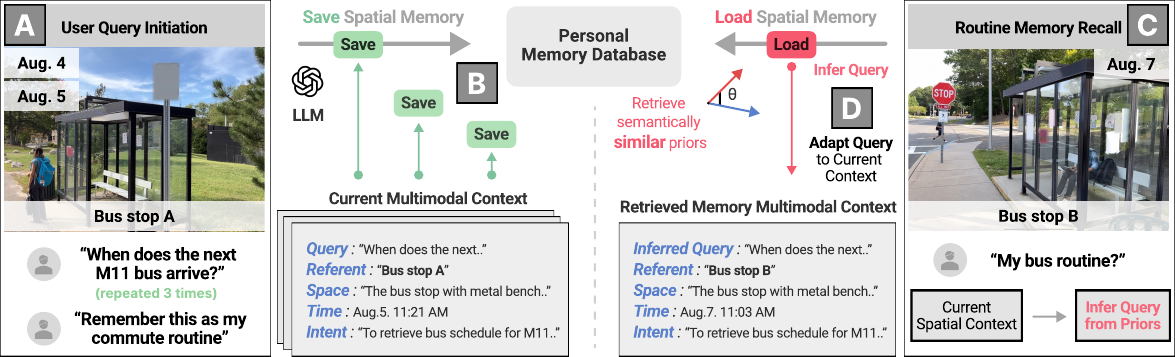}
    \vspace{-5mm}
    \caption{\sys pipeline overview: Repeated queries \textbf{(A)} labeled as ``commute routine'' are stored in a personal memory data-storage \textbf{(B)}, where a memory recall request \textbf{(C)} retrieves similar contextual memories, and uses them to infer the intention behind a user's new query, and proactively adapt to the user's current context \textbf{(D)}. The context of bus schedule and personal routine are extrapolated from memories}
    \label{fig:pipeline}
\vspace{-5mm}
\end{figure*}

\subsection{Design Goals and Considerations}
\label{subsec:design_goals}
Our longitudinal in-the-wild formative study (\cref{subsec:formative_study}) revealed several frictions in the everyday use of a wearable assistant. In the following section, we detail how each challenge motivated our design:

\vspace{1.75mm}
\recaptitle{\dc{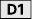} Socially Appropriate Intent Expression}
Fully articulated speech was found to limit the AI assistance in socially constrained environments, even when the assistance was desired. To address this, \sys is designed to decouple assistance from mandatory verbal articulation. Speech is treated as a flexible mechanism for expressing a user's needs (or intent), allowing the user to vary how explicitly they communicate (from full to micro/zero-utterance), depending on social context or situational convenience. We aim to mitigate the discomfort of public speech while preserving access to expressive interaction when needed.

\vspace{1.75mm}
\recaptitle{\dc{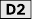} Spatially Grounded Personal Memories for Routines}
The informational needs of a wearable assistant were largely routine and repetitive, yet required users to repeatedly articulate the queries. We collect the user's personal context across everyday places and activities, identifying patterns in a long-term view. By binding user queries to personal multimodal context--space, time, action, intent, and referents--over time, we identify the plausible intent behind every user activity context~\cite{kim2025explainable, jang2005unified, lee2019remote}.

\vspace{1.75mm}
\recaptitle{\dc{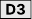} Proactivity with User Controlled Disambiguation}
While our spatial memories can proactively predict the intent of an action without explicit user specification, the robustness and transparency are key components to ensure a satisfactory AR assistance~\cite{xu2023xair}. Thus, we employ a human-in-the-loop, surfacing a confirmation interface (Yes/No) with rationale upon uncertainties (\cref{fig:confirmation_interface}).

\vspace{1.75mm}
\recaptitle{\dc{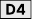} Minimal Interaction under Practical Constraints}
At the time of the study, the major hardware constraints of a wearable assistive device (Meta RayBan Gen-1), were battery drain (lasting 30 mins under consecutive video recording), overheating, and the social cost of verbal wake-word invocation (``\textit{Hey Meta}''). Therefore, we favor a minimal, on-demand interaction model that reduces continuous sensing and computation load--no SLAM or gesture recognizer necessitating continuous frame access. As a minimalistic fallback mechanism to speech-based interaction, we map four button-based triggers mirroring the Meta RayBan tap gestures (single, double, triple, and tap-and-hold) (\cref{fig:confirmation_interface}).

\section{Implementation of \sys}
\recaptitle{Interaction Paradigm: Speech as Granularity Control}
Speech is treated as the primary channel for explicit intent expression due to its ability to convey user needs and context within a single interaction. It is an effective mode of communication to express intentions, especially for limited or no-screen devices such as smart AR wearables~\cite{rajaram2025gesture}. In \sys, we leverage speech as a granularity control mechanism that provides users with direct agency over how explicitly their informational needs are specified. Users can selectively structure verbal queries to indicate intent, ranging from fully articulated queries with details, to partial core keyword-only utterances, and to zero-utterance interaction, depending on their situational context (\dc{figures/DC1.pdf}). To support this flexibility, we combine the general knowledge reasoning from LLM agents with long-term, personalized spatial memory (space, time, activity, referent), to extrapolate missing information/gaps in user intent specification (\dc{figures/DC2.pdf}). This design enables intent resolution even when verbal articulation is minimal or absent (\cref{fig:pipeline}). When agentic inference is uncertain, user feedback interface helps refine and confirm intent (\dc{figures/DC3.pdf}). We mirror the interaction mechanisms of commercially available smart wearable glasses (Meta RayBan), including verbal queries, memory remembrance, and minimal hardware-based gesture mappings (\dc{figures/DC4.pdf}), to assess the validity of our approach under practical conditions. We define three primary modes of our speech-based intent expression: \textit{Full}, \textit{Partial}, and \textit{Zero} Utterance:

\vspace{-3mm}
\begin{figure}[!h]
    \centering
    \includegraphics[width=\linewidth]{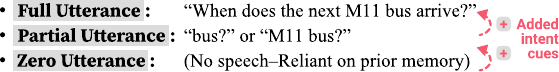}
    \vspace{-5.5mm}
    \caption{Varying granularity of speech inputs for intent expression.}
    \label{fig:paradigm_example}
\end{figure}
\vspace{-2mm}

\noindent
Micro-utterance interaction (\textit{Partial}, \textit{Zero}) extrapolates missing intent dimensions using the user's priors--spatial memory. Users can add cues (at their discretion) for more explicit intent expression (\cref{fig:paradigm_example}) (\textbf{even within \textit{Partial Utterance}}: ``bus'' to ``M11 bus'')\\ 
\vspace{-1mm}

\recaptitle{System overview}
\sys consists of six modules that jointly deliver context-aware intent resolution through speech granularity control and personal spatial memory: (1) \hyperref[subsec:query_decoding_agent]{Query Decoding Agent} interprets the user's interaction state by classifying the level of intent articulation and its type; (2) \hyperref[subsec:contextual_dim_encoder]{Contextual Dimension Encoder} constructs a structured ``dimension sketch'' that represents the user's ongoing spatial activity, combining verbal cues with embodied context; (3) \hyperref[subsec:personal_mem_store_manager]{Personal Memory Store Manager} 
incrementally builds long-term, personalized memories by binding prior interactions, inferred intents, and responses to spatiotemporal and referential contexts; (4) \hyperref[subsec:memory_retriever]{Memory Retriever} loads relevant past episodes with contextual similarity to the user's current state, enabling reuse and intent extrapolation; (5) \hyperref[subsec:response_composer]{Answer Composer} synthesizes concise responses by integrating the current scene with retrieved personal memories, and (6) \hyperref[subsec:user_verification_interace]{User Verification Interface} provides a light confirmation interface for uncertain intents. Note that the primary input channel of \sys is speech supplemented by buttons as a fallback mechanism to initiate a query or provide confirmations.
\vspace{-2mm}


\subsection{Query Decoding Agent}
\label{subsec:query_decoding_agent}
Query Decoding Agent serves as the entry point of the pipeline, ensuring that each incoming utterance (or zero-utterance) is routed appropriately. Every query is categorized into three types: \textit{Question Answering}, \textit{Remembrance}, and \textit{Removal}, decomposing the request into a set of sub-tasks. This guarantees reasoning transparency in the pipeline, and deterministic control over each step (over an autonomous decision making agentic design).

\parac{Question Answering}
represents a question-answering type. Within this category, we define three subtypes based on the level of verbal specification: \textit{`Full Utterance'}--a query that explicitly states the user intent, and can be answered directly from the current view or prior knowledge/memory; \textit{`Partial Utterance'}--under-specified or includes an implicit intent revealed only partially in the query, requiring intent inference drawn from prior user interactions; \textit{`Zero Utterance'}--zero-verbal cue, where the system fully infers the user's intent from spatial (visuo-location) memory contexts.

\parac{Remembrance}
applies when a user expresses intent in saving a memory (``\textit{Can you note that..}'' or ``\textit{Remember that..}'') without asking for a response (unidirectional command). The scene description, object-of-focus, time, and intent are encoded into an episodic entry and prepared for storage directly via \hyperref[subsec:personal_mem_store_manager]{Memory Store Manager}.

\parac{Removal}
provides users the freedom to manage their saved memories. Upon a verbal request (``\textit{Can you remove the memory on..}''), the system identifies the best-matching memory (semantically and keyword-based), summarizes it (for explainability), and requests user confirmation. Once confirmed, the memory is removed.

\subsection{Contextual Dimension Encoder}
\label{subsec:contextual_dim_encoder}
Dimension Encoder extracts the user's multimodal input into a compact ``dimension sketch.'' This sketch captures the user's current \textit{space label}, \textit{scene description}, \textit{referent/object-of-focus}, \textit{temporal moment}, \textit{action intent}, and the \textit{raw speech transcription}. These dimensions are bound jointly to form an instance of a ``spatial memory.'' For under-specified intents under Question Answering queries (\textit{Partial} or \textit{Zero}), the encoder passes these dimensions to the intent resolution step, where it attempts to refine and infer the intent of the query based on the user's stored/prior spatial memories. 

\subsection{Memory Retriever}
\label{subsec:memory_retriever}
Once a speech input is classified as Question Answering, and requires prior context of the user's interaction (\textit{Partial}, \textit{Zero}), the module computes contextual similarity between the current dimension sketch and stored episodes. It surfaces up to five relevant final candidates ($k=5$) to the user's intent and spatiotemporal context. The Retriever operates along: GPS proximity, semantic similarity (text, scene, and referent), and temporal recency/filter (if specified, e.g., time==``Tuesdays''); based on their semantic embeddings, re-ranking, and dimension fusion via Reciprocal Rank Fusion.

\parac{Static Source Prior Knowledge}
If a query involves simple recalling of a prior `as-is' memory, \sys recalls knowledge directly from \hyperref[subsec:personal_mem_store_manager]{Memory Store Manager}, combining retrieved episodes with the current scene. No external knowledge access (Web) is required, and responses are grounded in user's personal memories.

\parac{Live Source Prior Knowledge}
If the user refers to a memory that involves a live, up-to-date content (e.g., weather, transit), an updated response is needed. The Retriever proactively updates the response of the retrieved memory via the Google Custom Search API~\cite{googlecustomsearch}, before returning it as candidate memories.

\parac{Fresh Knowledge (No Prior)}
If a requested information in the user's query does not have any prior relevant memories (or has low confidence), but was requested with \textit{Partial} or \textit{Zero Utterance}, which requires query intent inference, the generic knowledge of the LLM is used to infer the intent. Then, it is validated by user via \hyperref[subsec:user_verification_interace]{User Verification Interface}, for a trustworthy response.

\subsection{Response Composer}
\label{subsec:response_composer}
\begin{figure}[!t]
    \centering
    \includegraphics[width=\linewidth]{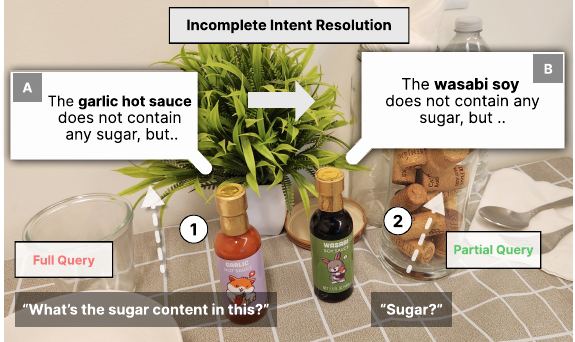}
    \vspace{-6.5mm}
    \caption{Proactive Intent Revision. \sys can adapt a query using spatial memories. A query asking the sugar content of \textbf{(A)} can proactively be adapted for \textbf{(B)} upon a shift in the referent-of-interest.}
    \label{fig:proactivity_explanation}
\vspace{-5.5mm}
\end{figure}

The Response Composer Agent follows a RAG-like pipeline~\cite{lewis2020retrieval}. It aggregates five candidate memories received from \hyperref[subsec:memory_retriever]{Retriever}, and together with the user query and current view, it generates an informed response. The agent is instructed to generate an adaptive response with respect to the user's current context (\cref{fig:proactivity_explanation}), with a concise answer (up to 30 words) to maintain simplicity for both visual and audio assistance. The composed answer is validated by an Answer Validation Agent that evaluates the response with a 1-10 confidence score. This response is transmitted to the client with a short rationale appended, following the principles of an explainable framework~\cite{xu2023xair}, ensuring that every reasoning step is accompanied by a rationale, guiding the user's understanding of ``\textit{why here}'' and ``\textit{why now}''. For a low confidence response, the user is asked to verify via \hyperref[subsec:user_verification_interace]{Verification Interface}.

\subsection{Personal Memory Store Manager}
\label{subsec:personal_mem_store_manager}
Personal Memory Store Manager serves as a long-term, per-user episodic memory layer. Each entry binds the query, its response, and the dimension sketch that contextualizes the interaction (\textit{scene description}, \textit{object}, \textit{time}, \textit{intent}, and \textit{spatial metadata}), in the form of a ``Spatial memory.'' This storage grows as part of the user's routines. Its design privileges routine-specific recalls, ensuring that personal memories can be re-accessed in future contexts. Furthermore, the store operation is gated by \hyperref[subsec:user_verification_interace]{User Verification}, with deletions supported at any time, upon confirmation. A low-confidence response triggers a user correction/verification rather than being silently recorded, to maintain high-quality memory storage. A memory is stored by either Remembrance or Question Answering.

\subsection{User Verification Interface and Visualization}
\label{subsec:user_verification_interace}
This ensures that the intention inference remains accountable and user-centric. Each low-confidence response is paired with a binary confirmation--accept or reject (Yes/No), along with its rationale (\cref{fig:confirmation_interface}). This human-in-the-loop mechanism allows users to curate their memory corpus with simplicity.

For visualization, query responses are presented either as screen-space UI (transcription of a verbal query) or as AR-anchored overlays (response). Users are given the option to toggle between the two modes (UI overlay vs AR situated text) allowing users to closely view distant AR content, while providing the option for a situated visual response. For the latter, we leverage LLM-based referent localization~\cite{gemini} to support open-vocabulary object detection. The identified referent is localized by combining visual and depth cues, enabling 3D placement of the AR content. This anchoring mechanism can act as a visual feedback that disambiguates the referent of the response, particularly when the user's query contains ambiguous references (under-specified verbal descriptions), where it is unclear which referent is associated with the response.


\section{Use cases and Applications}

\begin{figure*}[!t]
    \centering
    \includegraphics[width=\linewidth]{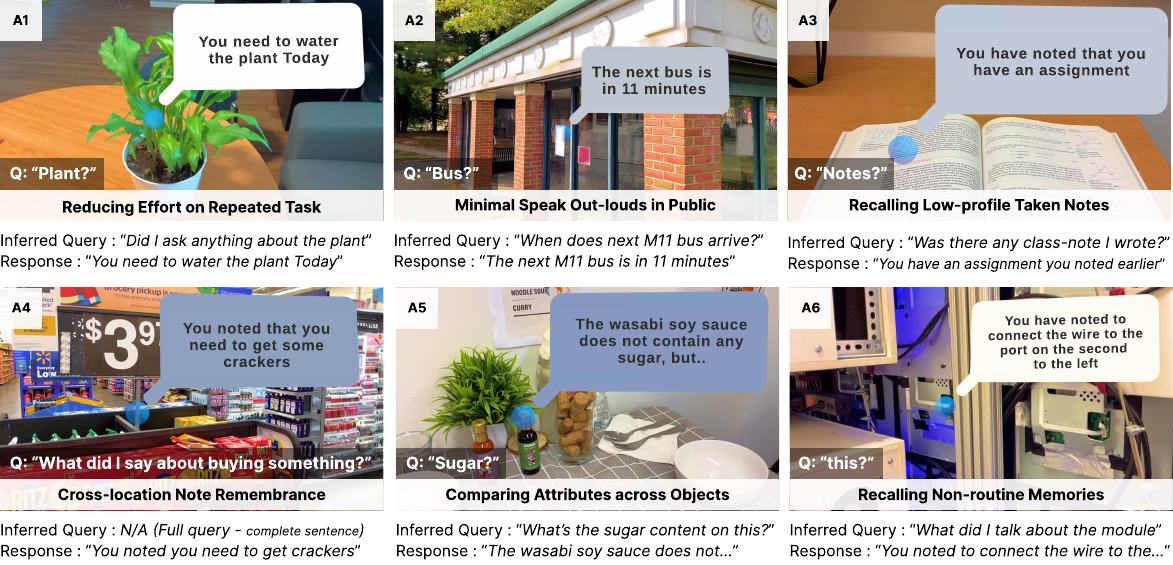}
    \vspace{-7.5mm}
    \caption{The six real-world use case scenarios of \sys. The scenarios were inspired by its ``in-the-wild'' usages in \hyperref[subsec:study_B]{Study B}.}
    \vspace{-2mm}
    \label{fig:usecase_and_applications}
\vspace{-3.5mm}
\end{figure*}

We illustrate how \sys is used, to reduce social awkwardness, maintain continuity of routines, and enable minimal or silent interaction during everyday activities. All visuals in \cref{fig:usecase_and_applications} were generated from in-app runs of \sys, denoted as \hyperref[fig:usecase_and_applications]{A1} to \hyperref[fig:usecase_and_applications]{A6}.

\vspace{-0.75mm}
\subsection{Home Routines: Reduced Efforts on Routine Tasks}
Household activities involve recurrent tasks that benefit from memory continuity. As shown in \hyperref[fig:usecase_and_applications]{A1}, a note ``\textit{Remind me to water the plant on Tuesdays}.'', can be simplified as ``\textit{plant?}.'' on the following Tuesday, while looking at the plant. \sys retrieves prior episodes associated with its temporal dimension, ``\textit{Tuesday}'' and its spatial (scene, referent, location) cues, producing a concise nudge: ``\textit{Plant watering on Tuesdays}.'' This shows how repetitive/routine queries can reduce users' verbal and remembrance effort.

\vspace{-0.75mm}
\subsection{Commute: Minimal Verbal Query among Crowd}
Commuting is a recurring routine where users require timely updates on transportation schedules or traffic conditions. At a frequently visited bus stop (\hyperref[fig:usecase_and_applications]{A2}), the user can obtain an up-to-date bus schedule without speaking aloud in public. Instead of fully articulating, ``\textit{When does the next M11 bus arrive?},'' ``\textit{M11 bus?}'' or ``\textit{routine bus schedule}'' (another memory-instance knowledge -- \hyperref[fig:pipeline]{\NoHyper\cref{fig:pipeline}\endNoHyper-A})) is sufficient. \sys proactively updates the live memory upon recall, ensuring that an accurate bus schedule is retrieved with minimal verbal cost in a socially sensitive context.

\vspace{-0.75mm}
\subsection{Classroom: Low-Profile Capture and Recall}
Speaking aloud in a classroom or a library is socially inappropriate. \sys enables low-profile interactions by supporting minimal utterances that are bound to the surrounding context. During a lecture, a student signals a ``Remembrance'' command while looking at course material, with a brief verbal note ``Assignment''. \sys records the visual description of the user's view, timestamp, and the referent (a chapter in a book) as part of the personal memory. Later, upon looking at the same course material, \sys can recall the assignment note anchored to the book (\hyperref[fig:usecase_and_applications]{A3}).

\vspace{-0.75mm}
\subsection{Grocery Shopping: Cross-space Information Recall}
Grocery lists frequently involve updating items, set elsewhere outside the store (\hyperref[fig:usecase_and_applications]{A4}). At home, a user may say, \textit{Remember to buy unsweetened soy milk from Walmart}.'' Later, at the store, when the user initiates a query: ``\textit{What did I say about buying something from Walmart?}'' Despite the spatial context disparity between home and store, \sys matches the intent of the query, retrieves the relevant memory, and surfaces the correct reminder.

\vspace{-0.75mm}
\subsection{Bistro: Within-space Cross-object Comparison}
A user's nutritional content information comparison can be achieved without repeatedly articulating lengthy queries (\hyperref[fig:usecase_and_applications]{A5}). A user initially asks, \textit{What is the sugar content of the wasabi soy sauce?},'' gazing at the target. \sys records the nutritional detail and the space as part of a memory. Later, while looking at a different sauce, \textit{sugar?}'' can recall the prior user interest at the Bistro, inferring the user's intent, and adaptively responds with the sugar content of the new sauce. This shows how \sys connects the experience even across objects while lowering effort.

\vspace{-0.75mm}
\subsection{Maintenance: Non-routine Memory Recall}
\sys is effective in recalling non-daily memories too (\hyperref[fig:usecase_and_applications]{A6}). During maintenance, a verbal note of ``\textit{Remember to connect the wire to the port second from the left}.'' left a month ago, can simply be glanced and uttered ``\textit{this}'' to recall the instruction notes. This is because the record of the spatial memory and its contextual cue is limited in number (uncommon visual features); retrieving the prior episode through \sys, requires even less interaction cost.


\begin{figure*}[!t]
    \centering
    \includegraphics[width=\linewidth]{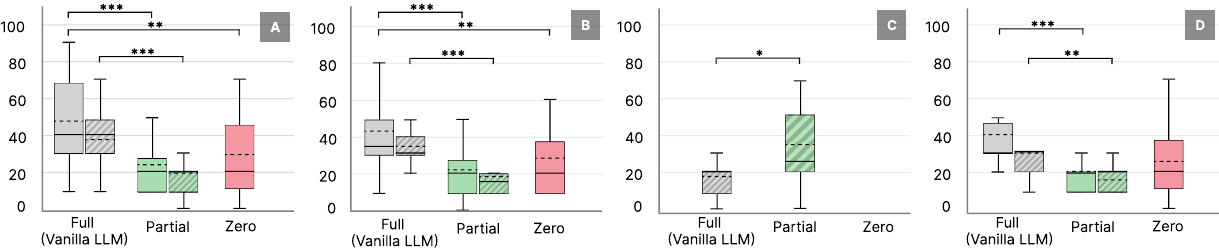}
    \vspace{-6.5mm}
    \caption{
    Comparison of participants' reported cognitive load with RTLX score : (A) Mental demand, (B) Physical demand, (C) Temporal demand, and (D) Effort; *** $p<.001$, ** $p<.01$, * $p<.05$; \textcolor{boxplot_gray}{\textbf{Full}},
    \textcolor{boxplot_green}{\textbf{Partial}},
    \textcolor{boxplot_red}{\textbf{Zero}} Utterance; 
    \includegraphics[height=0.75em]{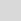} Task1 (Non-striped),
    \includegraphics[height=0.75em]{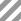} Task2 (Striped);
    }
    \label{fig:box_plot}
\vspace{-6mm}
\end{figure*}

\vspace{-0.5mm}
\section{Evaluation}
We evaluate \sys through two complementary studies. (1) \hyperref[subsec:study_A]{Study A}: a controlled lab study that assesses the benefits of spatially grounded memory and minimal-speech queries under counterbalanced conditions, and (2) \hyperref[subsec:study_B]{Study B}: an ``in-the-wild'' longitudinal deployment that examines the practical adoption of \sys in everyday settings assessing its social acceptability and privacy concerns. Together, we provide preliminary evidence on the social dynamics of a speech-triggered intent resolution mechanism, and how it adapts under varying spatial context, with the research questions:
\vspace{-5mm}

\begin{itemize}
\vspace{-1.5mm}
\item \para{RQ1} 
How does minimal/zero-utterance interaction influence human-AI interaction experience for daily routine activities?
\vspace{-2mm}
\item \para{RQ2} 
What does spatial memory mean for a personal wearable assistant? Is there any special advantages of it?
\vspace{-2mm}
\item \para{RQ3}
Does the minimal use of verbal queries shift user behaviors when using the AI assistant in public?
\end{itemize}
\vspace{-0.5mm}

We employed a within-subjects design with three utterance modes: {\textit{Full Utterance}} (denoted ``\fullq''), {\textit{Partial Utterance}}, and {\textit{Zero Utterance} (denoted ``\silentq''). The \fullq acts as our \uline{baseline}. Participants were informed of the data anonymization step, and provided consent prior to the participation. This study was conducted under the IRB approval of Stony Brook University (\textit{1173920}).

\vspace{-0.5mm}
\subsection{Study A: Understanding the Articulation Burden}
\label{subsec:study_A}
\vspace{-0.5mm}
We conduct technical and user analyses of \sys--accuracy, completion time, effort, perceived load, and usability under controlled lab setting. We concentrate on comparative analysis between vanilla AI assistance (\fullq) and micro (\partq, \silentq), to address \textbf{RQ1} and \textbf{RQ2}. In this study, we solely evaluate the \textit{\uline{interaction paradigm itself}--speech-based intent granularity control and articulation effort}, thus, we use a controlled experimental AR platform (HMD), to \textit{\uline{facilitate} precise assessment and to \uline{isolate our mechanism} from device ergonomics or form-factor effects}.

\cleansubsec{\textit{Study Design}}
\vspace{-1.75mm}
\para{Participants}
We recruited 18 participants (12 male, 6 female), aged 21-25 ($\mu=23.0$, $\sigma=1.19$) with prior experience with XR ($\mu=3.78$, $\sigma=2.02$; 7-point Likert scale). Participants reported Full Professional Proficiency or higher in English: Full Professional (N=14), Native/Bilingual (N=4). They were compensated \$15 for a 90-minute session. We denote them as P1 to P18. 

\para{Apparatus and Setup}
We conduct our experiment on a Meta Quest 3 HMD in passthrough AR mode with our custom Google Cloud Platform-hosted server logging user interactions. We use a controlled experimental platform (HMD) to capture the users' response in a deterministic, programmable, and reproducible manner (access to camera, parameters, hardware buttons). This reduces variability from external factors such as gesture-detection errors, ensuring that observed effects solely reflect the performance of our proposed interaction mechanism. We map the \hyperref[subsec:design_goals]{Minimalistic Tap Interface} to the Quest right controller `A' button due to limited access to its haptic touch input on the HMD.

We leverage the same Memory Retriever of \sys across all three conditions, as the current GPT API is limited in keeping the context within the conversation (``memory'') by default, requiring the API invoker to manually manage it. To showcase the adaptability of \sys under different space, time, intent, and objects, we made three personas, each indicating a hypothesized individual user, with disparate interests, and pre-seeded 45 personal memories before the study. This seed memory represents a hypothetical scenario where three distinct users have asked a number of questions prior to attempting to recall any memories. We construct three physical mock-up environments (Office, Break Room, and Bistro) to distribute these memories for a realistic use case scenario, and to study the adaptability of \sys in spatial memory recalls. Refer to \suppl for detailed study settings.

\para{Tasks}
Two controlled tasks were used to learn the interaction pattern of the participants with a conversational AI assistant across conditions, with referent grounding in realistic everyday contexts:

\parac{T1 - Remembrance} probes the recallability of previously seeded personal memories bound to objects and places (e.g., ``\textit{What did I say about $\{topic\}$}''). The task involves three physically different locations assessing spatial-memory aware information retrievals.

\parac{T2 - Comparison} probes information comparison between the object in the current view and another space or object (e.g., ``\textit{Can you compare the $\{topic\}$ between $\{obj1\}$ and $\{obj2\}$}''). The task evaluates an advanced use case of cross-space/object memory comparisons. \silentq was not included in this task due to lacking cues to extrapolate user intent.
\vspace{0.5mm}
  
\para{Procedure}
Each participant received a brief introduction of the study procedure followed by a 20-minute tutorial, familiarizing them with the HMD and the functionalities of \sys. The counterbalanced condition ordering and personas (45 seeded memories per persona) were given to the participant with the list of questions used during the memory seeding phase. The participants were given the freedom to randomly choose 18 (3 locations × 3+ trials × 2 task types) or more spatial memories across three different locations. Each participant was reminded to speak aloud and abide by the mode of interaction (\fullq, \partq, and \silentq). After each Task trial, they completed a user experience survey: 0-100 scaled RTLX (Raw NASA-TLX), 7-point Likert scale survey, SUS (System Usability Scale~\cite{sus}), and a semi-structured interview upon the completion of both Tasks. Repeated Measure ANOVA test (when normality held; Shapiro-Wilk test), or Friedman followed by Bonferroni corrected pairwise tests (or Wilcoxon signed-rank) was used.
\vspace{-4mm}

\cleansubsec{\textit{User Evaluation}}
\vspace{-1.5mm}
\para{Perceived Relevance}
Although no significance was identified across modes, participants reported that the answers were relevant to the spatial context. The relevance was measured with a 7-point scale (higher: more relevant). It showed that the mean score of \fullq was 5.94 ($\pm0.87$), followed by \partq, 5.44 ($\pm1.04$), and \silentq, 5.11 ($\pm1.68$). Overall spatial adaptability (being able to recall memory relevant to a space) was rated 5.78 ($\pm1.11$), indicating significance above the neutral score of 4 (Wilcoxon, $p<.001$) and robust perceived recall across the three environments.

\para{Cognitive Load}
We report the user's perceived load across six RTLX sub-dimensions: Mental, Physical, Temporal, Performance, Effort, and Frustration (higher: more load; 0-100 scaled). \cref{fig:box_plot} summarizes the distributions per condition and per significant pairs. Overall, abbreviating or removing articulation consistently reduced mental/physical/effort load without without harming perceived performance in T1, while in T2 the \partq condition lowered most loads but increased Temporal demand. No other dimensions showed statistical significance. 

In T1, the \textit{Mental demand} is ranked \fullq 47.78 ($\pm24.87$), \silentq 29.44 ($\pm27.96$), and \partq 25.00 ($\pm18.86$); (Friedman. ${\chi^2=23.41}$, ${p<.001}$); and for pairwise-test, \fullq$>$\partq (${p<.001}$), \fullq$>$\silentq (${p<.01}$). For both Physical demand and Effort, it follows the same trend. Physical demand: \fullq 42.78 ($\pm22.18$), \silentq 28.33 ($\pm23.33$), and \partq 22.22 ($\pm16.29$) with \fullq$>$\partq (${p<.001}$), \fullq$>$\silentq (${p<.01}$) (${\chi^2=19.00}$, ${p<.001}$). Effort: \fullq 40.00 ($\pm20.86$), \silentq 27.22 ($\pm21.91$), and \partq 21.11 ($\pm11.83$). However, in pair-wise analysis, only \fullq$>$\partq (${p<.001}$) was significant (${\chi^2=20.22}$, ${p<.001}$).

For T2, \partq reduced overall loads but increased temporal demand for some participants--without detailed keywords to allow \sys to pin-point referred memories, it required additional attempts to retrieve the desired response: ``\textit{\fullq query mode didn't take much but \partq, I had to wait longer to process my query sometimes.}'' (P10).
The Mental demand score is ranked: \fullq 37.22 ($\pm16.74$) and \partq 19.44 ($\pm12.59$), ($p<.001$). And the Physical demand indicates: \fullq 34.44 ($\pm10.42$) followed by \partq 18.33 ($\pm9.24$), ($p<.001$). The Temporal demand: \partq 34.44 ($\pm23.32$) higher than \fullq 18.89 ($\pm12.31$), ($p=.025$). Finally, the Effort: \fullq 32.22 ($\pm20.74$) and \partq 16.67 ($\pm6.86$), $p=.006$.

\para{Usability and Ease-of-use}
While participants reported \silentq as the most difficult approach to use (5.56$\pm1.95$) followed by \fullq (5.28$\pm1.74$) and \partq 4.78 ($\pm1.59$), the differences were not statistically significant across conditions. The mean perceived usability of the \sys interface (across all conditions) was rated `Good' with a SUS score of 75.42. 

The usefulness for reducing articulation burden in \partq was scored 5.39 ($\pm1.09$) and \silentq, 5.22 ($\pm1.48$), with significance compared to the neutral score of 4 (Wilcoxon, $p<.01$). Also, the use of \partq was evaluated easier to use compared to \silentq due to its controllability in providing direct cues (verbal) and transparency: 5.67 ($\pm1.08$); (higher: ease-of-use preference of \partq over \silentq) ($p<.001$); One of the participant said ``\textit{\silentq was the most difficult, because I can't control it to reason in a certain way. It made the task feel harder and more mentally demanding}'' (P17).

\para{User Feedback and Preferences}
In a questionnaire to rank the three conditions by preference (first being the most preferred), the result was evenly distributed among the conditions: \partq (N=7), \fullq (N=6), \silentq (N=5). Our interview with the participants revealed positivity towards using \partq and \silentq for convenience of use (P2, P5, P13), and \fullq for maximal explicitness (P1, P2, P4, P5, P13, P18). ``\textit{\fullq was more convenient for asking, \silentq was easy as nothing to speak. \partq conveyed intents but sometimes annoying, I had to think of the most effective keyword}''(P5); and ``\textit{\partq was helpful in reducing effort while capturing my intents. \silentq was great, I didn't have to say anything, it surprisingly knew.}'' (P13)
\vspace{-0.5mm}

\cleansubsec{\textit{Technical Evaluation}}
We collected 833 memories, each containing a dimension sketch, user query, and system response. We report latency, retrieval accuracy, and reduction in spoken effort.

\vspace{-0.25mm}
\para{Latency}
From a question to a response, the mean (in seconds) latency of \fullq was 3.52 ($\pm0.96 $)s followed by \partq 5.15 ($\pm1.38 $)s, and \silentq 2.42 ($\pm1.04 $)s. \silentq skips the transcription process as well as prior-based query intent analysis, directly inferring intent based on the visuo-spatial context, thus reaping a 31\% increase in speed over \fullq (aligning with user feedback). On the other hand, \partq extrapolates missing intent specifics using prior memories on top of question answering latency, resulting in a high-delay response.

\begin{figure}[!t]
    \centering
    \includegraphics[width=\linewidth]{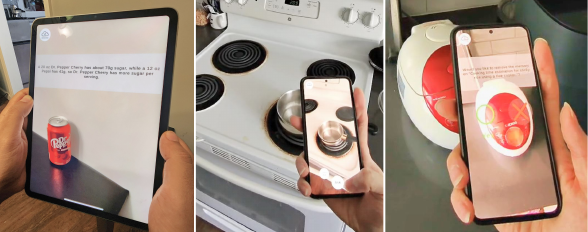}
    \vspace{-6.5mm}
    \caption{Field test of \sys on an established AR form factor.}
    \label{fig:user_study}
\vspace{-6.5mm}
\end{figure}

\para{Accuracy} 
The mean rate of successful QA for \fullq was 95.4\% ($\pm2.1$), while \partq inferred the intention behind the participants' query correctly, 86.7\% ($\pm3.4$) of the time. \silentq reaped 83.3\% ($\pm2.5$) accuracy. \silentq infers solely from the GPS location and visuals, not being able to capture intentions when there are more than one memory. The leading cause behind the \fullq errors was mispronounced questions, accent, or transcription failures.

\para{Measured Effort}
The average percentage decrease of word counts for \partq, compared to a \fullq query was ${49.8\%}$ ($\pm30.6$). This indicates a reduced user effort in attempting to initiate a memory-based query compared to vanilla speech-based interaction. We omit the comparison to \silentq, as it requires no spoken query.
\vspace{-0.5mm}

\subsection{Study B: Effects of \sys in Everyday Public}
\label{subsec:study_B}

This evaluation focuses on assessing the gains (or losses) that our \textit{interaction paradigm induces in \uline{social contexts}}. To examine system behavior under conditions where social discomfort is amplified, we conduct a longitudinal in-the-wild deployment. To evaluate the \textit{\uline{isolated effects}} of our speech-based interaction mechanism, we deploy \sys on a well-established smart device form factor (iOS, Android). Learned insights from \hyperref[subsec:formative_study]{Formative Study} indicate that smart-glasses form factors, as a newly introduced form factor, can \textit{draw social attention in public settings, \uline{even in the absence of any explicit verbal interactions}}. Such form factor-induced visibility introduces \textit{novelty and social-attention effects that are \uline{orthogonal to} the interaction mechanism of our evaluation}, limiting a fair assessment of the paradigm itself. By using mobile AR (socially familiar and widely accepted form factor), we \textit{\uline{minimize} these confounds} and \textit{\uline{directly evaluate} the system's ability to mitigate interaction friction} in everyday social environments (\textbf{RQ3}).

\cleansubsec{\textit{Study Design}}
\vspace{-1.5mm}
\para{Participants}
We recruited 13 participants (9 male, 4 female), aged 22-35 ($\mu=28.18$, $\sigma=4.61$) with prior experience with XR ($\mu=4.07$, $\sigma=2.17$; 7-point Likert). Participants reported their level of extroverted-ness ($\mu=4.01$, $\sigma=1.34$). They were compensated \$20 for a week-long participation. We denote them as P1 to P13. 

\para{System Setup}
The functionalities of \sys were maintained same as \hyperref[subsec:study_A]{Study A}, but ported to hand-held/mobile AR platforms in Unity. However, users were given two interfaces (software-level button) named ``Ask'' for \fullq and \partq queries, and ``Silent Ask'' for the \silentq query (\cref{fig:user_study}). Camera, Microphone, Network, and GPS sensors were accessed, and the app communicated with our Google Cloud Platform custom Python server.

\para{Tasks and Guidelines}
We maintained the task open-ended. Participants were instructed to ``speak aloud'' across diverse public and private settings, so that social comfort/awkwardness could be experienced. They were recommended to reach the daily quotas of 5 or more \fullq and 5 or more memory-based (\partq and \silentq) queries. Participants were told to perform tasks alone (no peers) to experience social pressures. We investigate the natural use of our interaction paradigm on AR: the guidelines and tutorials were provided, but the usages were not strictly mandated, and use the app as often as they can in their routine. A Google Form daily diary was provided to be filled at the end of the day. The participants were invited to a private Slack channel for any study inquiries.

\begin{table}[!t]
\centering
\caption{Qualitative coding of the ``in-the-wild'' diaries. The counts represent a comment (a comment can span across themes at once).}
\vspace{-1mm}
\label{tab:feedback_grouping}
\small
\begin{tabular}{|p{0.15\columnwidth}|p{0.725\columnwidth}|}
\hline
\textbf{Themes} & \textbf{Learnable Insights}\\
\hline
\textbf{Social\newline acceptability}\newline(N=18) & Default to \partq/\silentq in socially sensitive contexts. Only fall back to \fullq when confidence is very low or disambiguation is inevitably required. Helps in public attention mitigation.\\
\hline
\textbf{Public Privacy Concerns}\newline(N=7) & AR raising concerns due to a lack of established social-norms. People are unaware of what a new technology can do; I don't want to be stared at; The camera-like app in public drew more attention (misunderstood as taking a photo).\\
\hline
\textbf{Appropriate\newline for repeated tasks}\newline(N=14) & Realization of repeated use of smartphone during the study. Following similar pattern, daily, making use of the memory-based, repeating, with little effort. The app is simple. One interface for everything, no complicated conditions\\
\hline
\textbf{Limited Use cases}\newline(N=10) & Used the app for routine tasks such as weather checking, and temperature; Yet, suffered from limited use cases. May require longer-term use of the app.\\
\hline
\textbf{Queries with vague intent}\newline (N=19) & Ambiguous questions such as ``based on my prior questions'' or ``a new generic question'' leads to unexpected answers. Including a strong keyword such as ``my'' or ``remember that..'' helps clearly define intention.\\
\hline
\end{tabular}
\vspace{-6mm}
\end{table}

\cleansubsec{\textit{User Feedback and Experience}}
\vspace{-0.5mm}
We thematically group the participants' feedback from their diary (N=57)--(\cref{tab:feedback_grouping}), following the same procedure as \hyperref[subsec:formative_study]{Formative Study}. 

\para{Social Acceptability and Benefits}
Participants described \partq/\silentq as useful in social silence-defined or dense-crowd venues. One participant described the improved user experience: ``\textit{It's uncomfortable having to talk in the store. I'd rather use \silentq as much as possible until I need to speak to correct it.}'' (P7). Pressure amplified (e.g., Grocery store, Airport, Library) the benefit of minimal/zero articulation. ``\textit{Everyone was quiet. I got some stares when I spoke aloud. So \silentq was forced to be used}''(P5). In a 7-point Likert scale survey, micro/zero utterance substantially reduced the difficulty of ``speaking aloud in public when alone.'' The mean public speaking difficulty (low:better) of \fullq being 4.28 $(\pm1.82)$, micro interaction model was (\partq or \silentq), 2.63 $(\pm1.47)$.

\para{Usability and Preferences}
When asked to choose one mode for public/routine use, 63.2\% of diary responses preferred \partq/\silentq (N=36) over \fullq (N=21). The survey on the willingness to extend the capabilities of \silentq or \partq function on a smart wearable device (e.g., glasses) was 5.10 $(\pm1.81)$ (7-point scale), suggesting a positive tone with room for improvements. The primary reasoning behind the opposing view was due to the uncommonness of the form factor in society, ``\textit{people might stare at me}'' (P2) and its lack of differentiating use cases from mobile platforms (\cref{tab:feedback_grouping}). However, P2 indicated a strong need for a system such as \sys, underscoring the importance of verbal privacy-protection and its design for public use. A participant (P5) evaluated highly its value for daily reminders and repetitive tasks with reduced effort (\cref{tab:feedback_grouping}).

\para{User Feedback}
Overall, participants found micro-utterance useful. Especially for public scenarios, indicating positivity towards the angle of research. However, there were borderline cases where \hyperref[subsec:query_decoding_agent]{Query Decoding Agent} failed to parse the user's query intent between personal memory remembrance or a generic question inquiry. The natural language signals contain interpretation ambiguity (e.g., ``\textit{What is this?}''--can be a generic \fullq query, or may refer to user's personal meaning of the object), requiring nuanced, situational context to infer the query type. Another participant noted that the screen exposure of the mobile app to bystanders, increased public attention and reduced their comfortability when participating in public spaces such as a subway station. The participant requested a function that generates a `proxy' of the moment of query~\cite{satriadi2023proxsituated}, being able to concentrate on the content without having to worry about the AI assistant not being able to view the referent-of-question.
\vspace{-0.5mm}

\subsection{Takeaways and Analyses}
\vspace{-1mm}
Our studies showed that reducing articulation through \partq and \silentq lowered perceived cognitive and physical effort without substantially degrading accuracy, validating the benefit of minimal and implicit intent expression (\textbf{RQ1}). Furthermore, ``Spatial memory'' served as a personal context repository and effectively bridged under-specified or absent speech with environmental cues, enabling robust intent extrapolation across locations and activities (\textbf{RQ2}). In everyday public settings, participants naturally adopted our low-articulation modes in socially sensitive contexts, and resorted back to the explicit mode (\fullq) when full intent clarification was needed. This demonstrates that a user-centric speech-based intent granularity control mechanism can effectively mitigate social friction while preserving expressive capacity (\textbf{RQ3}). 

These findings derived across our \hyperref[subsec:formative_study]{Formative}, \hyperref[subsec:study_A]{Controlled}, and \hyperref[subsec:study_B]{Field} assessments, highlight the value of a per-situation intent resolution strategy that adapts how intent is expressed based on contextual constraints and task complexity. Rather than privileging a single interaction mode, \textit{we define the paradigm by treating speech as a flexible resource that users can regulate}, from \fullq articulation for explicit control, to \partq or \silentq utterance when reduced exposure or effort is desired. Also, we emphasize the practical importance of wearers' own privacy management and agency, beyond bystander-oriented privacy mechanisms (e.g., video recording visual/audio indicators), through our evaluation.

\vspace{-0.25mm}
\section{Limitation and Discussion}
\vspace{-1.75mm}
\para{Generality Across Wearable Form Factors}
We validated our proposed interaction paradigm (speech-based intent granularity control with spatial memory) rather than the full ergonomics and social experience of a smart glasses form factor. Although our proposal is not medium-specific, smart glasses might reap different outcomes than an HMD or mobile (our study setup), such as different comfort thresholds for speaking, or stronger bystander attention. Thus, we plan to extend \sys on smart glasses and replicate key measures under comparable routines to understand how form factor shifts the adoption boundary of the interaction paradigm.

\vspace{-0.05mm}
\para{Evolving Input Modalities Beyond Speech}
We frame speech as the most expressive and currently dominant channel for conveying complex intent on contemporary wearables. However, emerging techniques~\cite{rajaram2025gesture, lee2025sensible} and interfaces (e.g., Meta Neural Band~\cite{neuralband}) may change the interaction landscape. Our design goal is not to privilege speech exclusively, but to provide a general mechanism for regulating how explicitly intent is externalized while completing the remaining intent dimensions from personal context. In our future work, we can leverage new modalities as additional low-cost intent signals that \textit{complement} speech, while preserving the concept of the ``granularity control + contextual completion'' principle.

\vspace{-0.05mm}
\para{Recognizing Routine Patterns}
The routine activity detection of \sys currently relies on recency and word-based filters. However, the meaning of an action can shift substantially depending on its preceding context. More continuous modeling of user patterns is needed to capture everyday regularities. In future work, we will model repeated patterns beyond a single ``dimension sketch'' (space-time-referent-intent) by incorporating periodicity (daily/weekly) and semantic linking, enabling more nuanced retrieval and user needs prediction. We will also explore memory decay to mitigate long-term contextual overload~\cite{li2019combined}.

\vspace{-0.05mm}
\para{Screen-space vs Situated Interface}
We tackle the interaction paradigm for speech-controlled intent communication, a necessary step for wearable AR adoption, and allow users to have a manual choice between screen-space UI overlay and situated AR interface. However, practical wearable AR may require self-adaptive and context-awareness (e.g., topic-of-discussion, urgency, relevant referent, details, and occlusion~\cite{notificationurgency, song2025holistic, raianova2025adaptive, davari2020occlusion, davari2024towards, voxar}). We will extend our work towards an adaptive context-induced interface.

\vspace{-0.05mm}
\para{Arbitrating Multi-dimensional Contexts}
\sys resolves the fusion of spatiotemporal, intent, and verbal cues through Reciprocal Rank Fusion. While this method is functional, it lacks the robustness to handle nuanced conflicts, such as when a spoken reference contradicts a visual referent in view or when multiple dimensions suggest equally valid retrievals. Although designing an all-rounded recommender system is difficult, we plan to integrate dynamic inclusion/exclusion of dimensions by user's current context (parsing the needed ``dimension sketch'' using an agent), and add routine-pattern similarity cues.

\vspace{-0.05mm}
\para{Hardware Constraints and Sensing}
We highlighted that current wearable devices impose practical constraints that shape what interaction paradigms are viable in everyday use--short battery life under continuous sensing, overheating, and limited on-device compute for sustained perception and LLM inference. These constraints also interact with privacy: continuous capture raises bystander concerns and increases the need for selective logging and filtering of non-essential information. We adopted an on-demand interaction model, prioritizing lightweight intent resolution over continuous lifelogging. As smart wearable domain shifts rapidly, including hardware and OS advances (e.g., Android XR~\cite{androidxr}), we will revisit continuous, proactive information mining with the latest devices.

\vspace{-0.05mm}
\para{Towards Fully Proactive Wearable AR Assistance}
While \hyperref[subsec:contextual_dim_encoder]{Contextual Dimension Encoder} supports context-aware retrieval upon user initiation, we will extend \sys toward proactivity by modeling predictive intent from user preferences (e.g., OmniActions~\cite{omniactions}, Sensible Agent~\cite{lee2025sensible}) and personalizing with learning-based methods (e.g., RL, SFT) using episodic memories and feedback from \hyperref[subsec:user_verification_interace]{Verification Interface}. We also plan to reduce micro-utterance latency by precomputing candidate intents, benefiting from emerging lightweight real-time LLMs.
\vspace{-1mm}

\section{Conclusion}
\vspace{-1mm}
\sys introduced a proof-of-concept wearable AR assistant with a \textit{speech-based intent granularity control paradigm} grounded in \textit{personalized spatial memory}. By allowing users to regulate how explicitly intent is expressed (from full articulation to micro/zero-utterance), \sys reduced articulation burden while reasonably preserving accuracy, usability, and user-agency, extending assistance to socially sensitive everyday settings. Through longitudinal, in-the-wild, and lab assessments, we demonstrated that our approach minimizes articulation and cognitive/physical effort while adapting everyday and socially constrained environments. We position \sys as a foundational step toward scalable wearable interaction models that balance contextual intelligence, privacy, and expressive control, supporting long-term personalization and context-adaptive AR assistance~\cite{milgram1994taxonomy, lee2023design}.
\vspace{-1.5mm}

\acknowledgments
{
\vspace{-0.5mm}
We express our gratitude to Prantik Howlader, Nicholas Fraschilla-Brodkin, and Darius Coelho for their valuable feedback. This work was supported in part by NSF awards IIS2529207, IIS2441310 and ONR award N000142312124. \cref{fig:teaser} was edited using ChatGPT.
}

\bibliographystyle{abbrv-doi-hyperref-narrow}

\bibliography{references}
\end{document}